\documentclass[12pt]{iopart}

\usepackage{iopams}
\usepackage{graphicx}
\usepackage{graphics}
\usepackage{epsf}

\begin{document}
\title{Topological edge states in spin 1 bilinear-biquadratic model}

\author{Peng Li}
\address{Center for Theoretical Physics, Department of Physics, Sichuan University, Chengdu 610064, China}
\ead{lipeng@scu.edu.cn}

\author{Su-Peng Kou}
\address{Department of Physics, Beijing Normal University, Beijing 100875, China}
\ead{spkou@bnu.edu.cn}

\begin{abstract}
The spin $1$ bilinear-biquadratic model $H=\sum_{\left\langle ij\right\rangle }\left[  \cos\phi\mathbf{S}_{i}\cdot\mathbf{S}_{j}+\sin\phi(\mathbf{S}_{i}\cdot\mathbf{S}_{j})^{2}\right]$ 
on square lattice in the region $0<\phi<\pi/4$ is studied in a fermion representation with a $p$-wave
pairing BCS type mean-field theory. Our results show there may exist a non-trivial 
gapped spin liquid with time-reversal symmetry spontaneously breaking. 
This exotic state manifests its topological nature by forming chiral
states at the edges. To show it more clear, we set up and solved a ribbon system. 
We got a gapless dispersion representing the edge modes beneath the bulk modes. 
The edge modes with nonzero longitudinal momentum ($k_{x}\neq0$) convect 
in opposite directions at the two edges, which leads to a two-fold degeneracy. 
While the zero longitudinal momentum ($k_{x}=0$) modes turn out to be Majorana fermion
states. The edge spin correlation functions are found to decay in a power
law with the distance increasing. We also calculated the contribution of the
edge modes to the specific heat and obtained a linear law at low temperatures.
\end{abstract}

\maketitle

\section{Introduction}

In condensed matter physics, the Landau's theories of Fermi liquid and
spontaneous symmetry breaking have been the basic principles that account
for vast phenomena. For instance, the ground state of the two-dimensional
spin $S$ Heisenberg model on a square lattice possesses long range N\'{e}el
order and the Goldstone modes due to spontaneous spin-rotation symmetry
breaking. People have been trying to look for more exotic states in spin
systems for a long time. The seminal concept of \textquotedblleft resonating
valence bond\textquotedblright\ (RVB) spin liquid state was first proposed
by P. W. Anderson \cite{pwa}. Then, mostly due to its implication of the
mechanism of the high temperature superconductivity \cite{pwa2}, this field
has been flourishing for more than two decades. As a new type of quantum
matter, the spin liquid state itself is intriguing since its properties have
never been clarified before \cite{PALee}. Various approaches have shown that
quantum spin liquids may exist in two-dimensional ($2$D) $S=1/2$ \textrm{J}$%
_{1}$\textrm{-J}$_{2}$ model and the Heisenberg model on the Kagom\'{e}
lattice. In these models, the quantum spin liquids are accessed (in
principle) by appropriate frustrating interactions \cite{Fazekas}. However,
the nature of the quantum disordered ground state is still under debate. RVB
spin liquid state obviously goes beyond the Landau's theories, in which the
quasiparticles of Fermi liquid carry both spin and charge quantum numbers.
Another exciting field in searching for quantum exotic states beyond the
Landau's theory of spontaneous symmetry breaking is the quantum Hall (QH)
and fractional quantum Hall (FQH) states. In these states, topological order
plays an essential role \cite{2,qhe}. In the QH state, a quantized Hall
conductance was measured due to the formation of the Landau levels of a $2$D
electron gas at low temperatures and in strong magnetic field. One of the
key features of QH effect is the existence of chiral edge states around the
system boundaries. Recently, involving both quantized spin Hall effect \cite%
{kane,zhang} and quantized anomalous Hall effect \cite{haldane}, the
topological insulator with gapless edge states has been one of the hot
issues. The fundamental links between the above two fields have attracted
much attention \cite{wen}. In this work, we aim to contribute to these
interesting topics.

We study the spin $1$ bilinear-biquadratic model on square lattice, of which
the Hamiltonian is written as
\begin{equation}
H=\sum_{\left\langle ij\right\rangle }\left[ \cos \phi \mathbf{S}_{i}\cdot
\mathbf{S}_{j}+\sin \phi (\mathbf{S}_{i}\cdot \mathbf{S}_{j})^{2}\right] ,
\label{HBB}
\end{equation}%
where $\mathbf{S}_{i}$ is a spin $1$ operator. Its semiclassical version ($%
S\rightarrow \infty $) on a bipartitie lattice exhibits four ordered phases
that are exactly divided by four SU($3$) symmetric points at the model
parameters, $\phi =\phi _{0}=\pi /4,\pm \pi /2,-3\pi /4$ \cite%
{ChenHH,Papanicolaou}. They are phases with ferromagnetic (FM),
antiferromagnetic (AFM), ferroquadrupolar (FQ), and antiferroquadrupolar
(AFQ) orders respectively. Whether these classically well-understood phases
are stable in the quantum case or how the quantum model behaves constitues
the interesting topics of current researches. In one dimension ($1$D), many
aspects have been revealed by extensive exploration \cite%
{Haldane,Affleck,Fath,Fath2000,Xiang1993,Lauchli,Schollwock,Rommelse,Kennedy,White1993,Polizzi,Murashima,Gu}%
. While in two dimensions ($2$D), a complete understanding of the system is
still being anticipated. On triangular lattice, many methods have revealed
that the region $\pi /4<\phi \leq \pi /2$ exhibits an AFQ order \cite%
{Tsunetsugu,Lauchli2,Li2007}. On the honeycomb lattice, a tensor
renormalization group method showed the AFQ order in the region $\pi /4<\phi
\leq \pi /2$ is destroyed by pure quantum fluctuations and there is a
transition from the plaquette order to the AFM order \cite{Xiang2011}. A
recent work proposed that a three-sublattice order exists in the SU($3$)
point $\phi _{0}=\pi /4$ \cite{Tamas}. The quantum Monte Carlo simulation
\cite{Harada} found the AFM phase is stable in the region $-\pi /2<\phi \leq
0$ on a square lattice. While in the region of $0<\phi <\pi /4$, there lacks
evidence on whether the AFM order can survive or not.

In this paper, we show that a novel type of topological spin liquid might
exist on a square lattice in the region $0<\phi<\pi/4$, in which a
topological edge state circulates around the boundary of the system.
Time-reversal symmetry is broken spontaneously for this non-trivial gapped
spin liquid. In a fermion representation we found that the $2$D spin liquid
can be described very well by the (projected) spinless $p$-wave pairing
Bardeen-Cooper-Schrieffer (BCS)-type Hamiltonian. We obtained the gapless
dispersion of the edge modes. For a nonzero longitudinal momentum $%
k_{x}\neq0 $, one edge mode splits into two half modes that can exist
individually, which could be termed $\mathcal{L}$(left) and $\mathcal{R}$%
(right) chiral modes, respectively. While for $k_{x}=0$, zero edge modes
emerge, which turn out to be Majorana fermion states \cite%
{Kitaev,zhoubin2011}.

The paper is organized as follows. In Section II a brief introduction of the
fermion representation is presented. In Section III the mean-field theory is
introduced and the solutions of a gapped chiral spin liquid are obtained.
Then the corresponding edge state of a ribbon system is explored in Section
IV. And a summary is made in Section V.

\section{Fermion representation with hard-core constraint}

Firstly we introduce a fermion representation for quantum spin $1$. Each
spin has three eigenstates $\left\vert m_{j}\right\rangle $ of $S_{j}^{z}$
with the eigenvalues $m_{j}=-1,0,+1$. We introduce three fermionic operators
to generate three independent states ($\mathbf{i}=\sqrt{-1}$),
\begin{eqnarray}
f_{i,1}^{\dag }\left\vert 0\right\rangle  &=&\frac{\mathbf{i}}{\sqrt{2}}%
\left( \left\vert m_{i}=-1\right\rangle +\left\vert m_{i}=1\right\rangle
\right) , \\
f_{i,2}^{\dag }\left\vert 0\right\rangle  &=&\left\vert m_{i}=0\right\rangle
, \\
f_{i,3}^{\dag }\left\vert 0\right\rangle  &=&\frac{1}{\sqrt{2}}\left(
\left\vert m_{i}=-1\right\rangle -\left\vert m_{i}=1\right\rangle \right) .
\end{eqnarray}%
In terms of $f$ operators, the spin operators can be expressed as
\begin{eqnarray}
S_{i}^{x} &=&\mathbf{i}(f_{i,2}^{\dag }f_{i,1}-f_{i,1}^{\dag }f_{i,2}), \\
S_{i}^{y} &=&\mathbf{i}(f_{i,3}^{\dag }f_{i,2}-f_{i,2}^{\dag }f_{i,3}), \\
S_{i}^{z} &=&\mathbf{i}(f_{i,1}^{\dag }f_{i,3}-f_{i,3}^{\dag }f_{i,1}).
\end{eqnarray}%
To restore the Hilbert space of spin 1, the hard-core constraint at each
site must be imposed,%
\begin{equation}
\sum_{\mu =1}^{3}f_{i,\mu }^{\dag }f_{i,\mu }=1.  \label{single}
\end{equation}%
In this way the Hamiltonian Eq. (\ref{HBB}) is mapped to a frustrated SU($3$%
) fermion model \cite{Li2004}
\begin{equation}
H=-J_{1}\sum_{\left\langle ij\right\rangle }:F_{ji}^{\dag
}F_{ji}:-J_{2}\sum_{\left\langle ij\right\rangle }B_{ji}^{\dag
}B_{ji}+\sum_{i}\lambda _{i}\left( \sum\limits_{\mu }f_{i,\mu }^{\dag
}f_{i,\mu }-1\right) ,  \label{Hfermion}
\end{equation}%
where $J_{1}=\cos \phi >0$, $J_{2}=\cos \phi -\sin \phi >0$, $::$ denotes
normal ordering of operators, $\lambda _{i}$ are the Lagrangian multipliers,
and the bond operators are defined as%
\begin{equation}
F_{ji}=\sum_{\mu =1,2,3}f_{j,\mu }^{\dag }f_{i,\mu },B_{ji}=\sum_{\mu
=1,2,3}f_{j,\mu }f_{i,\mu }.  \label{bo}
\end{equation}%
A similar fermion representation could be found in a recent work \cite%
{Ng2010}.

\section{Gapped spin liquid with time-reversal symmetry breaking}

\subsection{Bond-operator mean-field theory}

To find the ground state properties of this spin $1$ system, we take the
mean field approximation by introducing two order parameters for the bond
operators $B_{ji}$ and $F_{ji}$,%
\begin{eqnarray}
\left\langle F_{i+x,i}\right\rangle  &=&\left\langle F_{i+y,i}\right\rangle
=F, \\
-\mathbf{i}\left\langle B_{i+x,i}\right\rangle  &=&B_{x}=Be^{\mathbf{i}\eta
_{x}},-\mathbf{i}\left\langle B_{i+y,i}\right\rangle =B_{y}=Be^{\mathbf{i}%
\eta _{y}},
\end{eqnarray}%
where $F$, $B$, $\eta _{x}$, and $\eta _{y}$ are real and to be determined
self-consistently. Here we have taken a unform phase factor in $F$, which
turns out to be negligible when the mean-field equations are established.
Two phase factors are kept for $B$ field, and we will see the final results
only rely on the phase difference $\Delta \eta =\eta _{y}-\eta _{x}$. Under
these prescription, the effective Hamiltonian reads%
\begin{eqnarray}
H_{eff} &=&\sum_{i,\mu }\lambda f_{i,\mu }^{\dag }f_{i,\mu
}+\sum_{\left\langle ij\right\rangle ,\mu }\left( f_{j,\mu }^{\dag
}T_{ji}f_{i,\mu }+h.c.\right) +\sum_{\left\langle ij\right\rangle ,\mu
}\left( f_{j,\mu }^{\dag }P_{ji}f_{i,\mu }^{\dag }+h.c.\right)   \nonumber \\
&&-\lambda N_{\Lambda }+2N_{\Lambda }J_{1}F^{2}+2N_{\Lambda }J_{2}B^{2},
\end{eqnarray}%
where $T_{ji}=-2J_{1}\left\langle F_{ji}\right\rangle $, $%
P_{ji}=-2J_{2}\left\langle B_{ji}\right\rangle $, and $N_{\Lambda }$\ is the
total number of lattice sites. The hard-core constraint shall be imposed on
an average level by minimizing the free energy. After performing the Fourier
transformation, we arrive at a complex $p$-wave-like pairing of independent
flavor of fermions \cite{Read},
\begin{eqnarray}
H_{eff} &=&\frac{1}{2}\sum_{\mathbf{k},\mu }\Phi _{\mu }^{\dag }(\mathbf{k}%
)M(\mathbf{k})\Phi _{\mu }(\mathbf{k})+\varepsilon _{0},  \label{Heff} \\
\varepsilon _{0} &=&\frac{1}{2}\lambda N_{\Lambda }+2N_{\Lambda
}J_{1}F^{2}+2N_{\Lambda }J_{2}B^{2}.
\end{eqnarray}%
where the sum of momentum $\mathbf{k}$ is carried out in the first Brillouin
zone ($1^{st}BZ$), the spinor $\Phi _{\mu }^{\dag }(\mathbf{k})=\left( f_{%
\mathbf{k,}\mu }^{\dag },f_{-\mathbf{k,}\mu }\right) $, the $2\times 2$
Hermitian matrix
\begin{equation}
M(\mathbf{k})=\mathbf{d}(\mathbf{k})\cdot \mathbf{\sigma }  \label{Mk}
\end{equation}%
with the Pauli matrices $\mathbf{\sigma }=(\sigma _{x},\sigma _{y},\sigma
_{z})$\textbf{\ }and%
\begin{eqnarray}
\mathbf{d}(\mathbf{k}) &=&\left( d_{x}(\mathbf{k}),d_{y}(\mathbf{k}),d_{z}(%
\mathbf{k})\right) ,  \label{dk} \\
d_{x}(\mathbf{k}) &=&2J_{2}B(\cos \eta _{x}\sin k_{x}+\cos \eta _{y}\sin
k_{y}), \\
d_{y}(\mathbf{k}) &=&2J_{2}B(\sin \eta _{x}\sin k_{x}+\sin \eta _{y}\sin
k_{y}), \\
d_{z}(\mathbf{k}) &=&\lambda -2J_{1}F\left( \cos k_{x}+\cos k_{y}\right) .
\end{eqnarray}%
The Lagrangian multiplier $\lambda _{i}$ is taken to be site-independent, $%
\lambda _{i}=\lambda $, which can also be regarded as a mean field. $%
N_{\Lambda }$ is the total number of lattice sites, and in fact we have
defined in this way a mean field Hamiltonian that is similar to the $2$D
Kiteav model for the $p+\mathbf{i}p$ superconductors for spinless fermions
\cite{zhoubin2011,Potter}. But the hard core constraint in Eq. (\ref{single}%
) may make it different. By performing the Bogoliubov transformation, one
can diagonalize the Hamiltonian as
\begin{equation}
H_{eff}=\frac{1}{2}\sum_{\mathbf{k},\mu }\Psi _{\mu }^{\dag }(\mathbf{k})P(%
\mathbf{k})\Psi _{\mu }(\mathbf{k})+\varepsilon _{0},
\end{equation}%
where $P(\mathbf{k})=\omega (\mathbf{k})\sigma _{z}$, $\Psi _{\mu }^{\dag }(%
\mathbf{k})=\left( \gamma _{\mathbf{k,}\mu }^{\dag },\gamma _{-\mathbf{k,}%
\mu }\right) $ with $\gamma _{\mathbf{k,}\mu }=u_{\mathbf{k}}f_{\mathbf{k,}%
\mu }-v_{\mathbf{k}}f_{-\mathbf{k,}\mu }^{\dag }$, where the coefficients
satisfy%
\begin{eqnarray}
\left\vert u_{\mathbf{k}}\right\vert ^{2} &=&\frac{1}{2}\left[ 1+\frac{d_{z}(%
\mathbf{k})}{\omega (\mathbf{k})}\right] , \\
\left\vert v_{\mathbf{k}}\right\vert ^{2} &=&\frac{1}{2}\left[ 1-\frac{d_{z}(%
\mathbf{k})}{\omega (\mathbf{k})}\right] , \\
2u_{\mathbf{k}}^{\ast }v_{\mathbf{k}} &=&\frac{d_{x}(\mathbf{k})+\mathbf{i}%
d_{y}(\mathbf{k})}{\omega (\mathbf{k})}.
\end{eqnarray}%
If one chooses a real and even $v_{\mathbf{k}}$ ($v_{\mathbf{k}}^{\ast }=v_{%
\mathbf{k}}$), then $u_{\mathbf{k}}$ is complex and odd, and vice versa. The
spectrum is
\begin{equation}
\omega (\mathbf{k})=\left\vert \mathbf{d}(\mathbf{k})\right\vert =\sqrt{%
d_{x}^{2}(\mathbf{k})+d_{y}^{2}(\mathbf{k})+d_{z}^{2}(\mathbf{k})}.
\label{spectrum}
\end{equation}%
The free energy can be worked out as%
\begin{equation}
F=-\frac{3}{\beta }\sum_{\mathbf{k}}\ln (1+e^{-\beta \omega (\mathbf{k}%
)})+E_{0},
\end{equation}%
where $\beta =\frac{1}{k_{B}T}$ and the ground state energy is%
\begin{equation}
E_{0}=-\frac{3}{2}\sum_{\mathbf{k}}\omega (\mathbf{k})+\varepsilon _{0}.
\label{E0}
\end{equation}%
By optimizing the free energy with respect to the mean fields, one obtains
the mean-field equations as the follows,
\begin{eqnarray}
\frac{1}{3} &=&\frac{1}{N_{\Lambda }}\sum_{\mathbf{k}}\frac{d_{z}(\mathbf{k})%
}{\omega (\mathbf{k})}\tanh \frac{\beta \omega (\mathbf{k})}{2},
\label{mfe1} \\
F &=&\frac{3}{2}\frac{1}{N_{\Lambda }}\sum_{\mathbf{k}}\frac{-d_{z}(\mathbf{k%
})\left( \cos k_{x}+\cos k_{y}\right) }{2\omega (\mathbf{k})}\tanh \frac{%
\beta \omega (\mathbf{k})}{2}, \\
B\left( \cos \eta _{x}+\cos \eta _{y}\right)  &=&\frac{3}{2}\frac{1}{%
N_{\Lambda }}\sum_{\mathbf{k}}\frac{d_{x}(\mathbf{k})\left( \sin k_{x}+\sin
k_{y}\right) }{\omega (\mathbf{k})}\tanh \frac{\beta \omega (\mathbf{k})}{2},
\\
B\left( \sin \eta _{x}+\sin \eta _{y}\right)  &=&\frac{3}{2}\frac{1}{%
N_{\Lambda }}\sum_{\mathbf{k}}\frac{d_{y}(\mathbf{k})\left( \sin k_{x}+\sin
k_{y}\right) }{\omega (\mathbf{k})}\tanh \frac{\beta \omega (\mathbf{k})}{2}.
\label{mfe4}
\end{eqnarray}%
All the mean fields can be determined by solving the set of mean-field
equations self-consistently. It is remarkable that all dispersions for
quasi-particles are three-fold degenerate at the mean field level.
Coexistence of non-zero solutions for both mean fields $F$ and $B$ affirms
the meaningful bond operator decomposition scheme in Eq. (\ref{Hfermion})-(%
\ref{bo}). Notice that the spectrum Eq. (\ref{spectrum}) is a function of
the phase difference $\left\vert \Delta \eta \right\vert =\left\vert \eta
_{y}-\eta _{x}\right\vert $, we can also take $\left\vert \Delta \eta
\right\vert $ as the optimizing parameter. It is clear to see that the
effective Hamiltonian Eq. (\ref{Heff}) preserves the time-reversal symmetry
when $\left\vert \Delta \eta \right\vert =0$ \cite{Li2004}, and does not
when $\left\vert \Delta \eta \right\vert \neq 0$. The equations are solved
at zero temperature to reveal the ground state properties. The non-zero
solution of mean fields $\lambda $, $F$ and $B$ for several choices of phase
difference $\left\vert \Delta \eta \right\vert $ are illustrated in Figure
1. At zero temperature, one can get a simple form for the ground state energy%
\begin{equation}
E_{0}=-2N_{\Lambda }J_{1}F^{2}-2N_{\Lambda }J_{2}B^{2},
\end{equation}%
where $\lambda $ is cancelled due to the substitution of the mean-field
equations Eq. (\ref{mfe1})-(\ref{mfe4}) in Eq. (\ref{E0}). The numerical
solution for the ground state energy $E_{0}$ and the gap of energy spectrum,
$\Delta _{gap}=\min \left( \omega \left( \mathbf{k}\right) \right) $, are
illustrated in Figure 2. We found that the lowest energy state can be
reached by choosing the phase difference $\left\vert \Delta \eta \right\vert
=\pi /2$. This solution is a $p$-wave paired gapped spin liquid and breaks
the time-reversal symmetry. The gapped spin liquid revealed here can be
classified by the $Z_{2}$ invariant \cite{Kou}. And by the $Z_{2}$ invariant
in the topological spin liquid state, we find a special fermion parity
pattern at high symmetry points in momentum space: even fermion parity at $%
\mathbf{k}=(\pi ,\pi ),$ $\mathbf{k}=(0,\pi ),$ and $\mathbf{k}=(\pi ,0)$
and odd fermion parity at $\mathbf{k}=(0,0)$ (please see detailed
calculations in Appendix). The physical regime of the spin liquid should not
exceed the Heisenberg point $\phi _{H}=0$, because the regime with $-\pi
/2<\phi \leq \phi _{H}=0$ exhibits an antiferromagnetic order \cite{Harada}.
The meaningful numerical solutions with nonzero $F$ and $B$ for $\left\vert
\Delta \eta \right\vert =\pi /2$ ceases near $\phi \gtrsim -0.08$ (see
Figure 2), a little less than $\phi _{H}=0$. The discrepancy can be ascribed
to the crudeness of the mean-field theory. At the SU($3$) point $\phi
_{0}=\pi /4$ \cite{Tamas}, our result shows that the gap closes along the
loop line $\mathbf{k}^{\ast }=(k_{x}^{\ast },k_{y}^{\ast }):$ $\cos
k_{x}^{\ast }+\cos k_{y}^{\ast }=1/A$ with $A=2.412513447$, where the
spectrum behaves linearly as $\omega \left( \mathbf{k}\right) \sim c\left(
\mathbf{k}^{\ast }\right) \left\vert \mathbf{k}-\mathbf{k}^{\ast
}\right\vert $ with anisotropic velocity $c\left( \mathbf{k}^{\ast }\right) $%
. The gaplessness on the loop line does not imply any order, so we get a
gapless spin liquid that is highly degenerate in thermodynamic limit.

\begin{figure}[tbp]
\begin{center}
\includegraphics[
height=4.1952in,
width=4.3915in
]{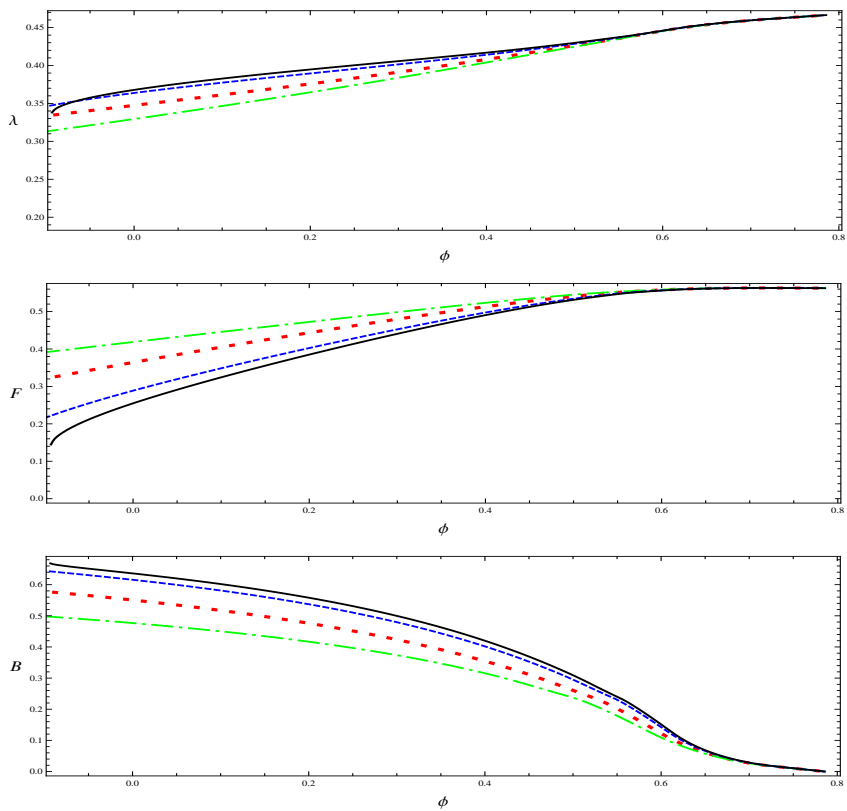}
\end{center}
\caption{(Color online) Numerical solution for the Lagrangian multiplier $%
\protect\lambda $ and the mean fields $F$ and $B$. The lines for several
selected phase differences are: $\left\vert \Delta \protect\eta \right\vert
=\left\vert \protect\eta _{x}-\protect\eta _{y}\right\vert =\protect\pi /2$,
black solid line; $\protect\pi /3$, blue dashed line; $\protect\pi /6$, red
dotted line; $0$, green dot-dashed line. Please see more details in the
text. }
\end{figure}

\begin{figure}[tbp]
\begin{center}
\includegraphics[
height=4.1961in,
width=4.3967in
]{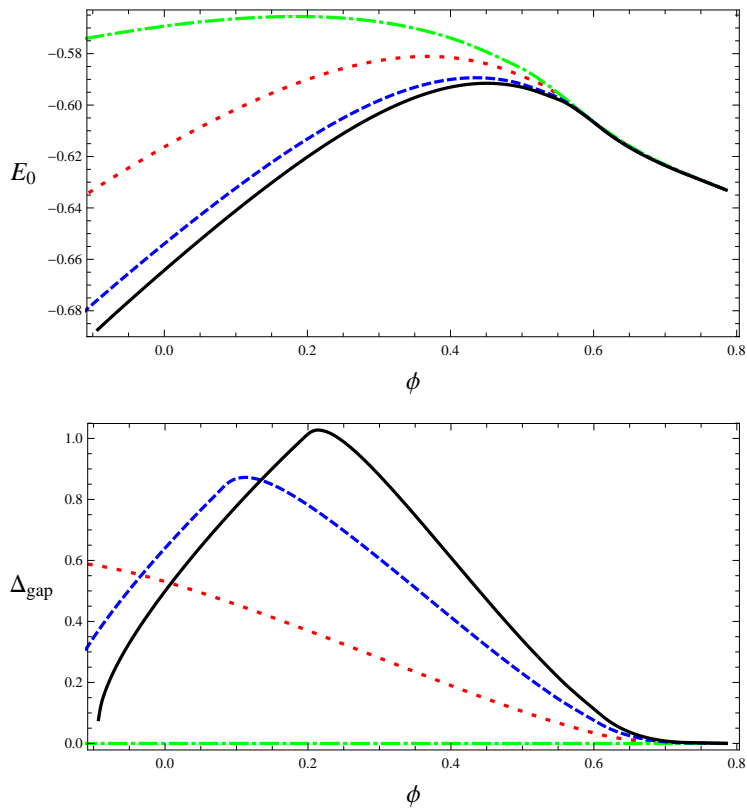}
\end{center}
\caption{(Color online) Upper: the ground state energy $E_{0}$; lower: the
gap of energy spectrum, $\Delta _{gap}=\min \left( \protect\omega \left(
\mathbf{k}\right) \right)$. The line types are the same as the ones
described in Figure 1.}
\end{figure}

\subsection{Ground state and the Chern number}

The $p$-wave paired ground state of the bulk system reads \cite{Read}%
\begin{equation}
\left\vert \Omega _{0}\right\rangle =\prod\limits_{\mathbf{k},\mu
}{}^{^{\prime }}\left( u_{\mathbf{k}}+v_{\mathbf{k}}f_{\mathbf{k,}\mu
}^{\dag }f_{-\mathbf{k,}\mu }^{\dag }\right) \left\vert 0\right\rangle ,
\label{groundstate0}
\end{equation}%
where the prime on the product indicates that each distinct pair $\left(
\mathbf{k},-\mathbf{k}\right) $ is to be taken once. This ground state
exhibits a non-trivial topological property that can be signified by the
Chern number of the spinless SU($3$) fermions. For each flavor of spinless
fermions, the Chern number is defined by \cite{Read,Cheng}
\begin{equation}
C=\frac{1}{4\pi }\int_{\Omega }d^{2}k[\mathbf{n}\left( \mathbf{k}\right)
\cdot \partial _{k_{x}}\mathbf{n}\left( \mathbf{k}\right) \times \partial
_{k_{y}}\mathbf{n}\left( \mathbf{k}\right) ]  \label{c}
\end{equation}%
where $\Omega $ means the volume of the first Brillouin zone, $\mathbf{n}%
\left( \mathbf{k}\right) $ is defined as $\mathbf{n}\left( \mathbf{k}\right)
=\frac{\mathbf{d}\left( \mathbf{k}\right) }{|\mathbf{d}\left( \mathbf{k}%
\right) |}$. By substituting Eq. (\ref{dk}) in Eq. (\ref{c}), we get%
\begin{equation}
C=\frac{\sin \left( \Delta \eta \right) }{4\pi }\left( \frac{2J_{2}B}{%
\lambda }\right) ^{2}\int_{\Omega }d^{2}k\frac{\left[ \frac{2J_{1}F}{\lambda
}\left( \cos k_{x}+\cos k_{y}\right) -\cos k_{x}\cos k_{y}\right] }{\left(
\omega \left( \mathbf{k}\right) /\lambda \right) ^{3}},
\end{equation}%
And by substituting the numerical mean-field solutions at zero temperature
in, we obtain the simplified result,
\begin{equation}
C=\pm 1,
\end{equation}%
for $0<\phi <\pi /4$. Thus the total Chern number of this topological state
is $C=\pm 3$ due to symmetry for the fermions of different flavors. The bulk
system's nontrivial ground state can be labeled by this Chern number.

\section{Spin edge states}

\subsection{Edge modes of the ribbon system}

To demonstrate the spin edge states explicitly, we set up a ribbon (or
ladders) system with a pair of open edges in $\hat{y}$ direction and keep
periodic boundary condition along $\widehat{x}$ axis (Figure 3). Noticing
that $k_{x}$ is still a good quantum number, we start with an Hamiltonian
with $L_{\max }$ legs,
\begin{eqnarray}
H_{eff}^{\prime } &=&\sum_{k_{x}\geq 0,\mu }\Phi _{\mu }^{\dag }\left(
k_{x}\right) M\left( k_{x}\right) \Phi _{\mu }\left( k_{x}\right)
+\varepsilon _{0}, \\
\Phi _{\mu }^{\dag }(k_{x}) &=&\left( f_{(k_{x},1),\mu }^{\dag
},...,f_{(k_{x},L_{\max }),\mu }^{\dag };f_{(-k_{x},1),\mu
},...,f_{(-k_{x},L_{\max }),\mu }\right) ,
\end{eqnarray}%
where we have parsed the zero momentum states in the first term and
restricted the sum to the positive values of momentum. The matrix $M\left(
k_{x}\right) $ are too large to be presented here. One can easily solve the
Hamiltonian numerically. The resulting diagonalized Hamiltonian could be
written in the form%
\begin{eqnarray}
H_{eff}^{\prime } &=&\sum_{k_{x}\geq 0,\mu }\Psi _{\mu }^{\dag }\left(
k_{x}\right) P\left( k_{x}\right) \Psi _{\mu }\left( k_{x}\right)
+\varepsilon _{0}, \\
P\left( k_{x}\right) &=&dia\left[ \omega _{\left( k_{x},1\right)
},...,\omega _{\left( k_{x},L_{\max }\right) };-\omega _{\left(
k_{x},1\right) },...,-\omega _{\left( k_{x},L_{\max }\right) }\right]
,(\omega _{\left( k_{x},i\right) }\geq 0), \\
\Psi _{\mu }^{\dag }(k_{x}) &=&\left( \gamma _{(k_{x},1),\mu }^{\dag
},...,\gamma _{(k_{x},L_{\max }),\mu }^{\dag };\gamma _{(-k_{x},1),\mu
},...,\gamma _{(-k_{x},L_{\max }),\mu }\right) ,
\end{eqnarray}%
or a further simplified one%
\begin{equation}
H_{eff}^{\prime }=\sum_{k_{x}\geq 0,\mu ,i}\omega _{\left( k_{x},i\right) }
\left[ \gamma _{(k_{x},i),\mu }^{\dag }\gamma _{(k_{x},i),\mu }+\gamma
_{(-k_{x},i),\mu }^{\dag }\gamma _{(-k_{x},i),\mu }\right] -\sum_{k_{x}\geq
0,\mu ,i}\omega _{\left( k_{x},i\right) }+\varepsilon _{0}.
\end{equation}

\begin{figure}[tbp]
\begin{center}
\includegraphics[
height=2.0in,
width=4.3261in
]{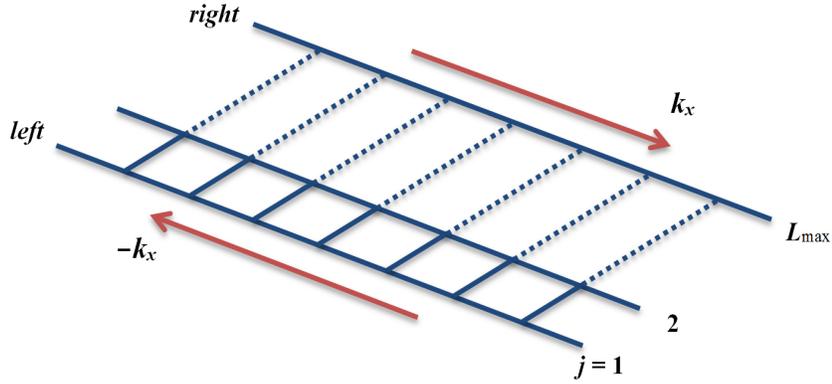}
\end{center}
\caption{(Color online) The ribbon system with open boundaries in $\widehat{y%
}$ direction. The edge modes at the two edges convect in opposite
directions. }
\end{figure}

\begin{figure}[tbp]
\begin{center}
\includegraphics[
height=2.9957in,
width=3.5769in
]{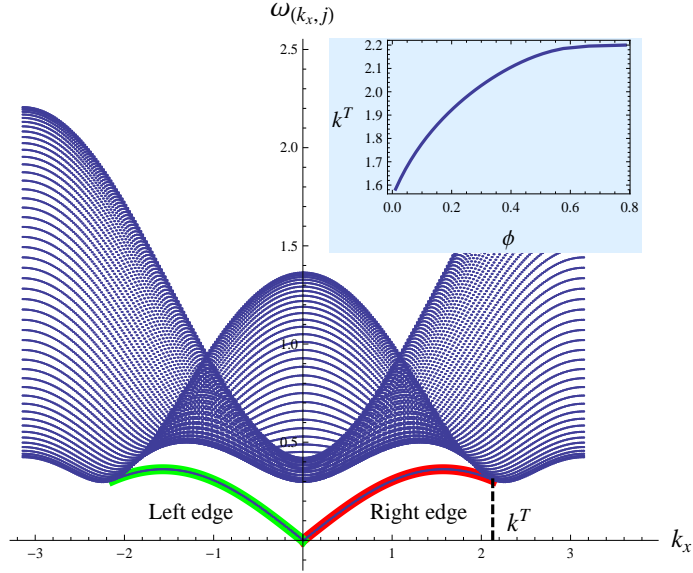}
\end{center}
\caption{(Color online) Numerical solutions of the spectra of the ribbon for
the selected model parameter $\protect\phi =0.435847$. The system's size is $%
L_{\max }=50$. The lowest thick line denotes the edge spectrum Eq. (\protect
\ref{edgespectrum}). Notice the gauge choice $\Delta \protect\eta =\protect%
\pi /2$ is taken here, so we have right modes for $k_{x}>0$ and left modes
for $k_{x}<0$. The inset shows the uppervalue $k^{T}$ for the edge states as
a function of model parameter $\protect\phi $.}
\end{figure}

We choose the canonical operator $\gamma_{(k_{x},1),\mu}^{\dag}$ (with
subscript $i=1$) to denote the edge excitations above the ground state and $%
\omega_{\left( k_{x},1\right) }$ the edge excitation energy. The rest modes
are bulk modes. The new ground state $\left\vert \Omega_{0}^{\prime
}\right\rangle $ is quite different from that in Eq. (\ref{groundstate0}). $%
\left\vert \Omega_{0}^{\prime}\right\rangle $ itself contains the chirality
of the edges and should satisfy the condition
\begin{equation}
\gamma_{(k_{x},i),\mu}\left\vert \Omega_{0}^{\prime}\right\rangle =0,\forall
k_{x},i,\mu.
\end{equation}
We will specify the ground state numerically later when evaluating some
quantities. The lowest energy mode for each $k_{x}$ could be collected as%
\begin{equation}
H_{lowest}^{\prime}=\sum_{k_{x}\geq0,\mu}\omega_{\left( k_{x},1\right) }
\left[ \gamma_{(k_{x},1),\mu}^{\dag}\gamma_{(k_{x},1),\mu}+%
\gamma_{(-k_{x},1),\mu}^{\dag}\gamma_{(-k_{x},1),\mu}\right] .
\end{equation}
It seems that this effective Hamiltonian denotes the edge states. However,
this is not necessarily the case. The lowest spectrum is a piecewise function%
\begin{equation}
\omega_{\left( k_{x},1\right) }=\left\{
\begin{array}{cc}
2J_{2}B\left\vert \sin k_{x}\right\vert , & (\left\vert k_{x}\right\vert
\leq k^{T}); \\
\sqrt{\left[ \lambda-2J_{1}F(1+\cos k_{x})\right] ^{2}+4J_{2}^{2}B^{2}%
\sin^{2}k_{x}}, & (\left\vert k_{x}\right\vert >k^{T}),%
\end{array}
\right.
\end{equation}
where $k^{T}=\arccos(\lambda/2J_{1}F-1)$ denotes a transition point. At this
transition point, both $\omega_{\left( k_{x},1\right) }$ and its first-order
derivative are continuous, but its second-order derivative is discontinuous.
One would find that only the first piece is the edge mode spectrum, i.e. the
edge mode spectrum ceases at $k^{T}$ and reads
\begin{equation}
\omega_{\left( k_{x},1\right) }^{edge}=2J_{2}B\left\vert \sin
k_{x}\right\vert ,(-k^{T}\leq k_{x}\leq k^{T}),  \label{edgespectrum}
\end{equation}
While the second piece is still a bulk mode spectrum. These conclusions are
testified by our numerical solutions.

\subsection{Numerical solution of the chiral edge states and zero mode
Majorana fermion states}

Now we discuss some more details about our numerical results. From the
numerical results, we confirm that the edge modes are well localized at the
edges. In practice, the solutions with $\Delta \eta =\pi /2$ and $\Delta
\eta =-\pi /2$ are degenerate, but have opposite chirality. So we only
demonstrate the solution for $\Delta \eta =\pi /2$ and $\phi =0.435847$ in
Figure 4. For $0<k_{x}\leq k^{T}$, we find the edge mode $\gamma
_{(-k_{x},1),\mu }^{\dag }$ localized at the left edge and $\gamma
_{(k_{x},1),\mu }^{\dag }$ at the right edge. Thus we may call them left ($%
\mathcal{L}$) and right ($\mathcal{R}$) chiral modes respectively,
\begin{eqnarray}
\mathcal{L}_{\mu }\left( -k_{x}\right) &\equiv &\gamma _{(-k_{x},1),\mu
}=\sum_{j=1}^{L_{\max }}\left[ U_{(-k_{x},j)}f_{(k_{x},j),\mu
}+V_{(-k_{x},j)}f_{(-k_{x},j),\mu }^{\dag }\right] , \\
\mathcal{R}_{\mu }\left( k_{x}\right) &\equiv &\gamma _{(k_{x},1),\mu
}=\sum_{j=1}^{L_{\max }}\left[ U_{(k_{x},j)}f_{(k_{x},j),\mu
}+V_{(k_{x},j)}f_{(-k_{x},j),\mu }^{\dag }\right] ,
\end{eqnarray}%
where the coefficients $U$ and $V$ are real (as is contrast to the
coefficients $u_{\mathbf{k}}$ and $v_{\mathbf{k}}$ for the periodic boundary
in previous section) and depicted in Figure 5(a) and (b). With the
increasing size of the system, the zero-momentum edge excitation $\omega
_{\left( 0,1\right) }^{edge}$ goes to zero rapidly. Beyond $L_{\max }=50$,
its energy value is so small that one can hardly decern it from the machine
precision. We may denote the zero modes as
\begin{equation}
\mathcal{E}_{\mu }\equiv \gamma _{(0,1),\mu }=\sum_{j=1}^{L_{\max }}\left[
U_{(0,j)}f_{(0,j),\mu }+V_{(0,j)}f_{(0,j),\mu }^{\dag }\right] ,
\end{equation}%
where the real coefficients are depicted in Figure 5(c). We see this zero
modes manifest itself at both edges. It is in fact a Majorana fermion state
\cite{Kitaev}, since the mode contributes zero energy to the system and can
be rewritten as
\begin{equation}
\mathcal{E}_{\mu }^{\dag }\mathcal{E}_{\mu }=1-2\mathbf{i}\mathcal{M}_{\mu
}^{\mathcal{L}}\mathcal{M}_{\mu }^{\mathcal{R}}
\end{equation}%
with two Majorana fermions%
\begin{eqnarray}
\mathcal{M}_{\mu }^{\mathcal{L}} &=&\sum_{j=1}^{L_{\max }}\frac{%
U_{(0,j)}-V_{(0,j)}}{2}\left( -\mathbf{i}f_{(0,j),\mu }+\mathbf{i}%
f_{(0,j),\mu }^{\dag }\right) , \\
\mathcal{M}_{\mu }^{\mathcal{R}} &=&\sum_{j=1}^{L_{\max }}\frac{%
U_{(0,j)}+V_{(0,j)}}{2}\left( f_{(0,j),\mu }+f_{(0,j),\mu }^{\dag }\right) ,
\end{eqnarray}%
localizing at the two opposite edges. All of the above coefficients satisfy
the relations numerically for large enough $L_{\max }$ $(k_{x}\geq 0)$%
\begin{eqnarray}
\sum_{j=1}^{L_{\max }}\left[ U_{(\mp k_{x},j)}^{2}+V_{(\mp k_{x},j)}^{2}%
\right] &=&1, \\
U_{(k_{x},j)} &=&V_{(k_{x},j)},U_{(-k_{x},j)}=-V_{(-k_{x},j)}, \\
U_{(-k_{x},j)} &=&-U_{(k_{x},L_{\max
}-j+1)},V_{(-k_{x},j)}=V_{(-k_{x},L_{\max }-j+1)}, \\
U_{(-k_{x},j)}U_{(k_{x},j)} &=&V_{(-k_{x},j)}V_{(k_{x},j)}=0.
\end{eqnarray}%
But notice each mode possess a U($1$) symmetry, so that the values may be
changed according to the symmetry transformation.

\begin{figure}[tbp]
\begin{center}
\includegraphics[
height=5.2589in,
width=4.8473in
]{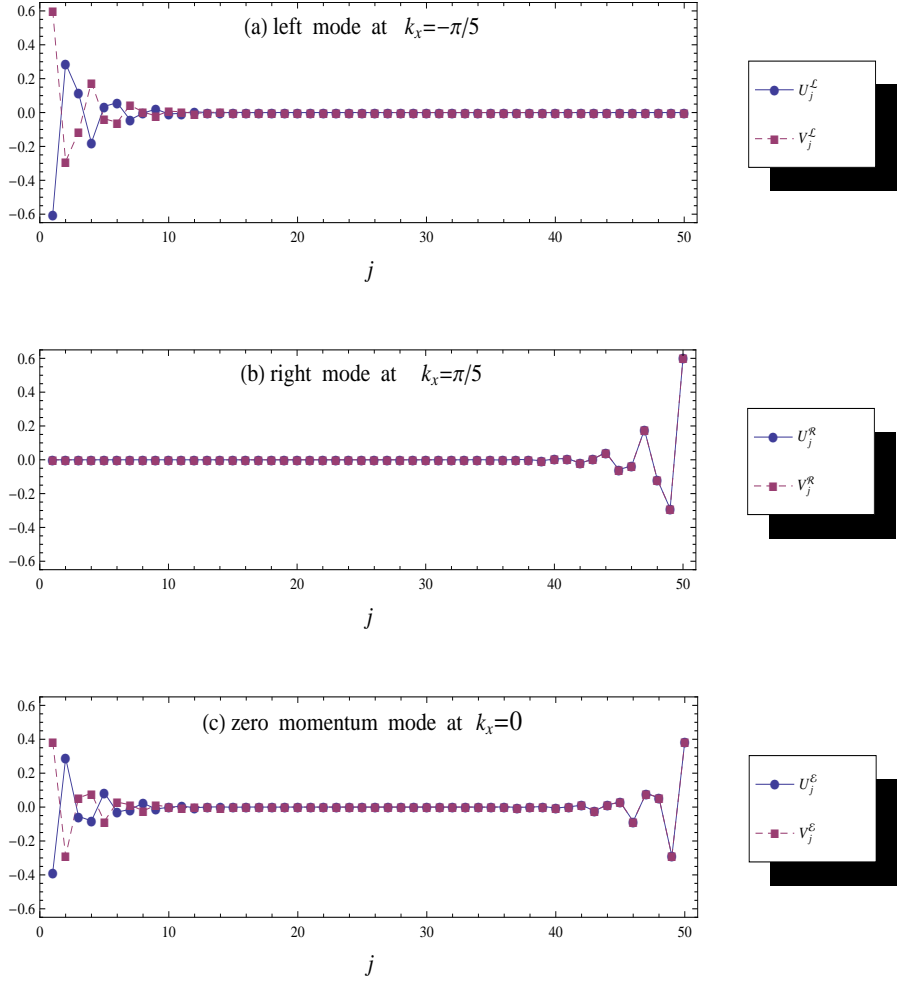}
\end{center}
\caption{Visulization of the edge modes in fermion representation on a
ribbon with width $L_{\max }=50$ ($1\leq j\leq L_{\max }$): (a) the left
modes at $k_{x}=-\protect\pi /5$, (b) the right mode at $k_{x}=\protect\pi %
/5 $ and (c) the zero momentum mode at $k_{x}=0$. The model parameter $%
\protect\phi =0.435847$ and gauge choice $\Delta \protect\eta =\protect\pi %
/2 $.}
\end{figure}

\subsection{Edge spin correlation functions}

Although the edge states of the ribbon are clear to see in the fermion
representation, it is still illusive from the point of view of the spin
language. In order to show the properties of the spin edge state, we measure
the spin correlations and thermodynamic quantities, such as the specific
heat, contributed by the edge.

It is well-known a gapped spin system exhibits an exponentially decaying
spin correlation in the bulk. Of all the spin correlations for the ribbon
system, the one at the edge is of our great interest. We choose the right
edge of the ribbon (Figure 3) to measure the spin correlations in the ground
state,%
\begin{equation}
C_{edge}^{zz}(i,i+r)\equiv C_{edge}^{zz}(r)=\left\langle S_{(i,L_{\max
})}^{z}S_{(i+r,L_{\max })}^{z}\right\rangle .  \label{corr}
\end{equation}%
Now we need to find out the the ground state $\left\vert \Omega _{0}^{\prime
}\right\rangle $. The edge modes in $H_{edge}^{\prime }$ could be singled
out and serve as a quasi-$1$D effective Hamiltonian,%
\begin{eqnarray}
H_{eff}^{\prime } &=&H_{bulk}^{\prime }+H_{edge}^{\prime },  \label{Hedge} \\
H_{edge}^{\prime } &=&\sum_{0\leq k_{x}\leq k^{T},\mu }\omega _{\left(
k_{x},1\right) }^{edge}\left[ \gamma _{(k_{x},1),\mu }^{\dag }\gamma
_{(k_{x},1),\mu }+\gamma _{(-k_{x},1),\mu }^{\dag }\gamma _{(-k_{x},1),\mu }%
\right] ,
\end{eqnarray}%
which could be utilized to evaluate quantities along the edges. Near $%
k_{x}\sim 0$, the edge modes behave linearly $\omega _{\left( k_{x},1\right)
}^{edge}\sim 2J_{2}B\left\vert k_{x}\right\vert $ \cite{zhoubin2011}(please
see the lowest thick line in Figure 4). Since the bulk and edge modes are
independent, one can write the ground state in a separable form
\begin{equation}
\left\vert \Omega _{0}^{\prime }\right\rangle =\left\vert \Omega
_{0,bulk}^{\prime }\right\rangle \otimes \left\vert \Omega _{0,edge}^{\prime
}\right\rangle ,
\end{equation}%
where $\left\vert \Omega _{0,bulk}^{\prime }\right\rangle $ and $\left\vert
\Omega _{0,edge}^{\prime }\right\rangle $ are the lowest energy states of $%
H_{bulk}^{\prime }$ and $H_{edge}^{\prime }$ respectively. We have%
\begin{equation}
\left\vert \Omega _{0,edge}^{\prime }\right\rangle ={\prod\limits_{0\leq
k_{x}\leq k^{T},j,\mu }}\left[ U_{(k_{x},j)}-V_{(k_{x},j)}f_{(k_{x},j),\mu
}^{\dag }f_{(-k_{x},j),\mu }^{\dag }\right] \left\vert 0\right\rangle ,
\label{gsedge}
\end{equation}%
where the prime on the product indicates that each distinct pair $\left(
k_{x},-k_{x}\right) $ is to be taken once. One can easily verify that $%
\gamma _{(\mp k_{x},1),\mu }\left\vert \Omega _{0,edge}^{\prime
}\right\rangle =0$. Since we are concerning the quantities along the edge,
the edge correlation function Eq. (\ref{corr}) can be evaluated
approximately just by $\left\vert \Omega _{0,edge}^{\prime }\right\rangle $.
At first, one can work out%
\begin{eqnarray}
\left\langle f_{(k_{x},L_{\max }),\mu }^{\dag }f_{(-k_{x},L_{\max }),\mu
}^{\dag }\right\rangle &=&-U_{(k_{x},L_{\max })}V_{(k_{x},L_{\max })},
\nonumber \\
\left\langle f_{(k_{x},L_{\max }),\mu }^{\dag }f_{(k_{x},L_{\max }),\mu
}\right\rangle &=&V_{(k_{x},L_{\max })}^{2}.
\end{eqnarray}%
Then the correlation function is deduced as%
\begin{eqnarray}
C_{edge}^{zz}(r) &=&2\left\vert \frac{1}{N_{\Lambda }}\sum_{0\leq k_{x}\leq
k^{T}}e^{\mathbf{i}k_{x}r}\left\langle f_{(k_{x},L_{\max }),\mu }^{\dag
}f_{(-k_{x},L_{\max }),\mu }^{\dag }\right\rangle \right\vert ^{2}  \nonumber
\\
&&+2\left\vert \frac{1}{N_{\Lambda }}\sum_{0\leq k_{x}\leq k^{T}}e^{\mathbf{i%
}k_{x}r}\left\langle f_{(k_{x},L_{\max }),\mu }^{\dag }f_{(k_{x},L_{\max
}),\mu }\right\rangle \right\vert ^{2}  \nonumber \\
&=&2\left\vert \frac{1}{N_{\Lambda }}\sum_{0\leq k_{x}\leq k^{T}}e^{\mathbf{i%
}k_{x}r}U_{(k_{x},L_{\max })}V_{(k_{x},L_{\max })}\right\vert
^{2}+2\left\vert \frac{1}{N_{\Lambda }}\sum_{0\leq k_{x}\leq k^{T}}e^{%
\mathbf{i}k_{x}r}V_{(k_{x},L_{\max })}^{2}\right\vert ^{2}.
\end{eqnarray}%
The results show a general power law
\begin{equation}
C_{edge}^{zz}(r)\approx \frac{\alpha }{r^{\delta }}.  \label{Cedge}
\end{equation}%
In practice, we fit the numerical data by the formula%
\begin{equation}
\ln C_{edge}^{zz}(r)\approx \ln \alpha -\delta \ln r
\end{equation}%
instead. At the model parameter $\phi =0.435847$, we obtain $\alpha \approx
0.0156407$ and $\delta \approx 2.174343$ (see Figure 6(a)). For other model
parameters, the results are not much different (Figure 6(b) and (c)). So we
see that the edge spin correlations decay like a power law along the edge
and exclude the possibility of exponential decay behavior in the main region
of model parameters.

\begin{figure}[tbp]
\begin{center}
\includegraphics[
height=4.4317in,
width=3.9369in
]{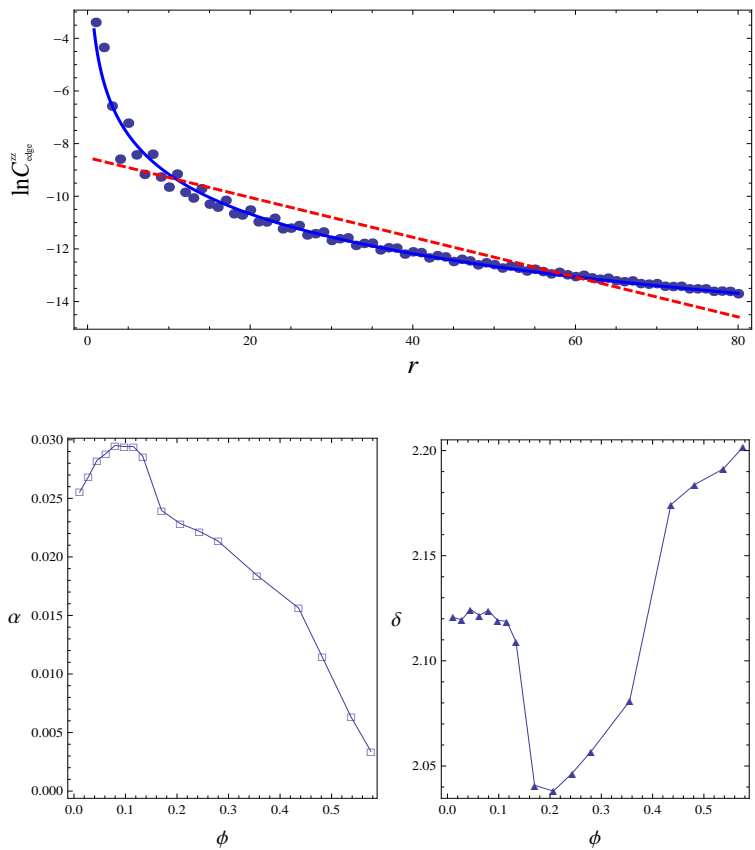}
\end{center}
\caption{(Color online) (a) The spin correlation functions (in logarithm)
along the right edge of the ribbon. The dots are the numerical results. Here
the width of the system is $L_{\max }=50$, the model parameter is $\protect%
\phi =0.435847$ and the gauge choice is $\Delta \protect\eta =\protect\pi /2$%
. The solid blue line is a power-law fit ($\protect\alpha \approx 0.0156407$
and $\protect\delta \approx 2.174343$). The red dashed straight line is an
exponential law fit. (b) The fitted coefficient $\protect\alpha $ as a
function of $\protect\phi $. (c) The fitted exponent $\protect\delta $ as a
function of $\protect\delta $. Please see more details in the text.}
\end{figure}

\subsection{Specific heat contributed by the edge modes}

Now we turn to the specific heat. At low temperatures, the bulk states
contribute little to the specific heat due to the existence of bulk gap,
while the gapless edge modes give the main contribution. By adopting Eq. (%
\ref{Hedge}) as the effective Hamiltonian, one can worked out the
contribution of the edge modes to the specific heat that behaves linearly in
temperature $T$,
\begin{equation}
\frac{C_{V}}{N_{\Lambda }k_{B}}=\int_{0}^{E_{m}}\left( \frac{E}{2k_{B}T}%
\right) ^{2}\cosh ^{-2}\left( \frac{E}{2k_{B}T}\right) \rho \left( E\right)
dE\approx \frac{\pi k_{B}}{8J_{2}B}T,
\end{equation}%
where we have released the upper limit of the integral for simplicity ($%
E_{m}\rightarrow \infty $) and the density of states is
\begin{equation}
\rho \left( E\right) =\frac{1}{N_{\Lambda }}\sum_{k_{x},\mu }\delta \left(
E-\omega _{(k_{x},1)}\right) \approx \frac{3}{2\pi J_{2}B}.
\end{equation}

\section{Summary}

In a brief summary, we showed a possible gapped chiral spin liquid in the $2$%
D square bilinear-biquadratic system in the region of $0<\phi<\pi/4$. As a
consequence, the time-reversal symmetry breaks spontaneously and an
interesting topological ground state is revealed. We numerically analysed
the resulting spin edge states for a ribbon system in detail. This method
may be applied to other relating systems to specify a spin liquid state. We
found $\mathcal{L}$(left) and $\mathcal{R}$(right) chiral edge modes for
nonzero longitudinal momentum $k_{x}\neq0$ and a zero edge modes for $%
k_{x}=0 $. The power-law decay of the edge spin correlation function and the
contribution of the nontrivial spin edge state to the specific heat at low
temperatures are found. In the future work, the properties of low energy
excitations would be of great interest.

\section*{Acknowledgement}

The authors thanks Professor Shun-Qing Shen for fruitful discussions. This
work was supported by SRF for ROCS SEM (20111139-10-2), the Chinese National
Natural Science Foundation under Grant No.: 11074177, 11174035. This
research was supported in part by the Project of Knowledge Innovation
Program (PKIP) of Chinese Academy of Sciences, Grant No. KJCX2.YW.W10.

\section*{Appendix: Classification topological state by $Z_{2}$ topological
invariants}

For each flavor of fermions (omit the flavor index $\mu $ in Eq. (\ref{Heff}%
)), the effective Hamiltonian is
\begin{equation}
H_{eff}=\frac{1}{2}\sum_{\mathbf{k}}\Phi ^{\dag }(\mathbf{k})M(\mathbf{k}%
)\Phi (\mathbf{k}),  \label{Hmean1}
\end{equation}%
with $M(\mathbf{k})$ defined in Eq. (\ref{Mk}). From the results in Ref.\cite%
{Kou}, the $2\times 2$ Pauli matrices can be divided into two groups - even
matrix $\sigma _{z}$ and odd matrices, $\sigma _{x}$ and $\sigma _{y}$.
Because the coefficients of odd matrices are zero at four high symmetry
points of square lattice in momentum space, we can only focus on the
coefficients of even matrix, $d_{z}(\mathbf{k})=\lambda -2J_{1}F\left( \cos
k_{x}+\cos k_{y}\right) $. The four $Z_{2}$ topological invariants are
defined as
\begin{equation}
\mathcal{\zeta }_{k}=1-\Theta (d_{z}(\mathbf{k})),
\end{equation}%
where $\Theta (x)=1$ if $x>0$ and $\Theta (x)=0$ if $x<0$.

Hence, for points $(0,0)$, $(0,\pi )$, $(\pi ,0)$, $(\pi ,\pi )$, the four $%
Z_{2}$ topological invariants are explicitly given by
\begin{eqnarray}
\mathcal{\zeta }_{k=(0,0)} &=&\Theta \lbrack \lambda -4J_{1}F],  \nonumber \\
\mathcal{\zeta }_{k=(0,\pi )} &=&\Theta \lbrack \lambda ],  \nonumber \\
\mathcal{\zeta }_{k=(\pi ,0)} &=&\Theta \lbrack \lambda ],  \nonumber \\
\mathcal{\zeta }_{k=(\pi ,\pi )} &=&\Theta \lbrack \lambda +2J_{1}F].
\end{eqnarray}%
For $\mathbf{k=}(\pi ,\pi ),$ $k=(0,\pi )$, $k=(\pi ,0),$ we have a trivial
result as
\begin{equation}
\mathcal{\zeta }_{\mathbf{k}=(\pi ,\pi )}=\mathcal{\zeta }_{k=(0,\pi )}=%
\mathcal{\zeta }_{k=(\pi ,0)}=0;
\end{equation}%
for $\mathbf{k=}(0,0),$ we have
\begin{equation}
\mathcal{\zeta }_{\mathbf{k}=(0,0)}=\Theta \left( \lambda -4J_{1}F\right) .
\end{equation}%
Thus we identify two distinct topological states: the topological state with
trivial topological invariants
\begin{equation}
\mathcal{\zeta }_{\mathbf{k}=(0,0)}=\mathcal{\zeta }_{\mathbf{k}=(\pi ,\pi
)}=\mathcal{\zeta }_{\mathbf{k}=(0,\pi )}=\mathcal{\zeta }_{\mathbf{k}=(\pi
,0)}=0
\end{equation}%
for $\lambda >4J_{1}F$ and the topological state with nontrivial topological
invariants
\begin{equation}
\mathcal{\zeta }_{\mathbf{k}=(\pi ,\pi )}=\mathcal{\zeta }_{\mathbf{k}%
=(0,\pi )}=\mathcal{\zeta }_{\mathbf{k}=(\pi ,0)}=0,\mathcal{\zeta }_{%
\mathbf{k}=(0,0)}=1
\end{equation}%
for $\lambda <4J_{1}F$. And in the topological spin liquid state in this
paper, we find a special fermion parity pattern: even fermion parity at $%
\mathbf{k}=(\pi ,\pi ),$ $\mathbf{k}=(0,\pi ),$ and $\mathbf{k}=(\pi ,0)$
and odd fermion parity at $\mathbf{k}=(0,0)$.

\end{document}